\documentclass[twoside]{article}
\usepackage{amsmath,amsfonts,amssymb}
\usepackage{algorithm}
\usepackage{algpseudocode}
\newcommand{\ket}[1]{|#1\rangle}
\newtheorem{property}{Property}
\newtheorem{lemma}{Lemma}
\newtheorem{corollary}{Corollary}
\newtheorem{theorem}{Theorem}
\newcommand{\Endproof}{\hfill$\Box$\\}
\newcommand{\Beginproof}{\noindent{\bf Proof:} }

\begin{document}
\setlength{\textheight}{8.0truein}    


\setcounter{page}{1}


\vspace*{0.88truein}



\centerline{\bf
QUANTUM ALGORITHMS FOR THE SHORTEST COMMON SUPERSTRING }
\centerline{\bf
AND TEXT ASSEMBLING PROBLEMS}
\vspace*{0.37truein}
\centerline{\footnotesize
KAMIL KHADIEV}
\vspace*{0.015truein}
\centerline{\footnotesize\it Institute of Computational Mathematics and Information Technologies, Kazan Federal University}
\baselineskip=10pt
\centerline{\footnotesize\it Kremlyovskaya, 35, Kazan, Russia}
\vspace*{10pt}
\centerline{\footnotesize 
CARLOS MANUEL BOSCH MACHADO}
\vspace*{0.015truein}
\centerline{\footnotesize\it Institute of Computational Mathematics and Information Technologies, Kazan Federal University, }
\baselineskip=10pt
\centerline{\footnotesize\it Kremlyovskaya, 35, Kazan, Russia}
\vspace*{10pt}
\centerline{\footnotesize 
ZEYU CHEN}
\vspace*{0.015truein}
\centerline{\footnotesize\it School of Mathematics Sciences, Zhejiang University,}
\baselineskip=10pt
\centerline{\footnotesize\it  Hangzhou 310058, P. R. China}
\vspace*{10pt}
\centerline{\footnotesize 
JUNDE WU}
\vspace*{0.015truein}
\centerline{\footnotesize\it School of Mathematics Sciences, Zhejiang University,}
\baselineskip=10pt
\centerline{\footnotesize\it  Hangzhou 310058, P. R. China}
\vspace*{0.225truein}

\vspace*{0.21truein}

In this paper, we consider two versions of the Text Assembling problem. We are given a sequence of strings $s^1,\dots,s^n$ of total length $L$ that is a dictionary, and a string $t$ of length $m$ that is texts. The first version of the problem is assembling $t$ from the dictionary. The second version is the ``Shortest Superstring Problem''(SSP) or the ``Shortest Common Superstring Problem''(SCS). In this case, $t$ is not given, and we should construct the shortest string (we call it superstring) that contains each string from the given sequence as a substring.
These problems are connected with the sequence assembly method for reconstructing a long DNA sequence from small fragments. For both problems, we suggest new quantum algorithms that work better than their classical counterparts. In the first case,  we present a quantum algorithm with $O(m+\log m\sqrt{nL})$ running time. In the case of SSP, we present a quantum algorithm with running time $O(n^3 1.728^n +L +\sqrt{L}n^{1.5}+\sqrt{L}n\log^2L\log^2n)$. 

\vspace*{10pt}

\vspace*{3pt}

\section{Introduction}        
\label{sec:intro}  
In this paper, we are interested in running-time-efficient solutions for two problems that are  The Shortest Common Superstring Problem and The Text Assembling Problem. In the general case, the problem is as follows. For a positive integer $n$, a sequence of $n$ strings $S=(s^1,\dots,s^n)$ is given. We call it a dictionary. We assume that the total length of the dictionary strings is $L=|s^1|+\dots+|s^n|$. Additionally, a string $t$ of length $m=|t|$ is given. We call it text. Our goal is to assemble $t$ from the dictionary strings $S$. Here we have two types of the problem:
\begin{itemize}   
    \item \textbf{The Shortest Common Superstring Problem (SCS).} It is also known as the ``Shortest Superstring Problem''(SSP). In this case, the text $t$ is not given. We should construct the shortest string $t$ (we also call it superstring) that contains each string from the dictionary $S$ as a substring.
     \item \textbf{The Text Assembling Problem.} The string $t$ is given, and we should assemble $t$ only using strings from $S$. Here, we can use a string from dictionary $S$ several times or not use it at all. We allow overlapping of dictionary strings during the assembly process.
\end{itemize}

These problems are connected with the sequence assembly method for reconstructing a long DNA sequence from small fragments \cite{msdd2000} which is a well-known problem in bioinformatics. The sequence assemble problem has two types. The first one is the Reference-guided genome assembly method that constructs an existing long DNA string from the sequence $S$. For the problem, we should know the string $t$ apriori and check whether we can construct it from $S$. This case is close to The Text Assembling Problem. The second type of the sequence assemble problem is de-novo assembly; in this problem, we do not have the string $t$ at all, and we should construct it using all strings from $S$. The Shortest Superstring Problem is used as one of the main tools for solving de-novo assembly problems \cite{bdfhw95}. The problem has applications in other areas such as virology and immunology (the SCS models the compression of viral genome); the SCS can be used to achieve data compression; in scheduling (solutions can be used to schedule operations in machines with coordinated starting times), and others. According to \cite{m78,m94,m98,v2005}, The Shortest Common Superstring Problem is NP-hard. So, approximation algorithms are also explored, the best-known algorithm is \cite{ks2013}. At the same time, researchers are interested in exact solutions also. The algorithm based on \cite{b62,hk62} has $O^*(2^n+nL)$ running time. Here $O^*$ notation hides polynomial factors. If we have a restriction on the length of the strings $s^i$, then there are better algorithms. If a length $|s^i|\leq 3$, then there is an algorithm \cite{gkm2013} with running time $O^*(1.443^n)$. For a constant $c$, if a length $|s^i|\leq c$, then there is a randomized algorithm \cite{gkm2014} with running time $O^*(2^{(1-f(c))n})$ where $f(c)= 1/(1 + 2c^2)$. We can see that $2^{(1-f(c))n}\geq 1.851^n$ for any $c\geq 2$.

The Text Assembling Problem is much easier.
It was considered in \cite{kr2021a}. Authors of that paper presented a deterministic algorithm with $O^*(m+L)$ and a lower bound is $\Omega(m+L)$. Here $O^*$ hides logarithmic factors. The other version of the problem that does not allow overlapping of string on assembling process \cite{kr2021b} has a randomized algorithm with running time $O^*(m\sqrt{L}+L)$ and a lower bound is also $\Omega(m+L)$.

We refer to \cite{nc2010,a2017,k2022lecturenotes,aazksw2019part1} for a good introduction to quantum algorithms. There are many problems where quantum algorithms outperform the best-known classical algorithms. Some of them can be founded here \cite{dw2001,quantumzoo}. 
Problems for strings are examples of such problems \cite{ki2019,kk2021,rv2003,l2020,l2020conf,aj2021,m2017,aaksv2022,ke2022}. One of the most popular performance metrics for quantum algorithms is \emph{query complexity}. So, we explore problems from this point of view.

The best-known quantum algorithm for The Text Assembling Problem \cite{kr2021a} has a running time 
$O\left(m +\log m\cdot(\log n+\log\log m)\cdot \sqrt{n\cdot L}\right).$
Note, that in the non-overlapping case \cite{kr2021b}, it is $O^*(mL^{1/4}+ \sqrt{nL})$. The quantum lower bounds for both cases \cite{kr2021a,kr2021b} are $\Omega(\sqrt{m}+\sqrt{L})$.
A quantum algorithm for SCS is not known at the moment.

In this paper, we present new quantum algorithms for these two problems:
\begin{itemize}
   
    \item \textbf{Shortest Common Superstring Problem(SCS).} We present a quantum algorithm for the SCS problem with running time 
    $O\left(n^{3}1.728^n+L+n^{1.5}\sqrt{L}+n\sqrt{L}\log^2L\log^2n\right).$ The algorithm is based on Grover’s search algorithm \cite{g96,bbht98}, Maximum search algorithm \cite{dh96,dhhm2006} and the Dynamic programming approach for a Boolean cube \cite{abikpv2019,b62,hk62}. Additionally, we used a quantum algorithm based on quantum string matching algorithm \cite{rv2003} for searching duplicates in $S$, and checking whether a string is a substring of one another string from $S$. We have this subproblem because strings in $S$ can have different lengths.  
    As far as we know, our algorithm is the first quantum algorithm for the SCS problem.

     \item \textbf{The Text Assembling Problem} We present a quantum algorithm with $O(m+\log m\cdot\sqrt{nL})$ running time. The algorithm is a modification of the algorithm from \cite{kr2021a} and uses a technique from \cite{k2014,lmrss2011}. The new idea is $(\log n+\log\log m)$ times faster than the known algorithm in the case the dictionary $S$ has a big size. It achieves a quantum speed-up if $O(n)$ strings from $S$ have length at least $|s^i|=\omega(\log^2 m)$, and $m=O(\log m\cdot\sqrt{nL})$. The condition is better than it was in \cite{kr2021a} and allows us to obtain a quantum speed-up in much more real-world cases.
\end{itemize}

The structure of this paper is the following. Section \ref{sec:prelims} contains preliminaries. We discuss the SCS problem in Section \ref{sec:csc}, and the Text Assembling problem in Section \ref{sec:tao}. Section \ref{sec:conclusion} concludes the paper.

The paper is the extended version of the paper \cite{kb2022} that was presented at International Conference on Micro- and Nano-Electronics 2021.

\section{Preliminaries}\label{sec:prelims}

Let us consider a string $u=(u_1,\dots,u_m)$. Let $u[i,j]$ denote a substring $(u_i,\dots,u_j)$ for $1\leq i\leq j\leq m$. Let $|u|=m$ be a length of a string $u$. We assume that $u_i\in \Sigma$ where $\Sigma$ is some finite size alphabet.
For simplicity, we assume that $\Sigma=\{0,1\}$, but all results are valid for any finite-size alphabet.

\subsection{Formal Definitions for Problems}
Let us discuss formal definitions of the problems.

\subsubsection{The Shortest Common Superstring Problem} For a positive integer $n$, a sequence of $n$ strings $S=(s^1,\dots,s^n)$ is given. We should construct the shortest string $t$ (we call it superstring), i.e. $|t|$ is the minimal possible such that each $s^i$ is a substring of $t$ for $i\in\{1,\dots,n\}$. In other words, for each $i\in\{1,\dots,n\}$ there is $1\leq q_i\leq |t|$ such that $t[q_i,q_i+|s^i|-1]=s^i$.  

Informally, we want to construct the shortest $t$ that contains all strings from $S$ as substrings. Let us denote the problem as $\texttt{SCS}(S)$.

\subsubsection{The Text Assembling Problem.}
For some positive integers $n$ and $m$,  a sequence of $n$ strings $S=(s^1,\dots,s^n)$ is given. We call $S$ a dictionary. Additionally, we have a string $t$ of length $m$. We call $t$ a text. We should present a sequence $s^{i_1},\dots,s^{i_r}$ and positions $q_1,\dots,q_r$ such that $q_1=1$, $q_r=n-|s^{i_r}|+1$, $q_j\leq q_{j-1}+|s^{i_{j-1}}|$  for $j\in\{2,\dots,r\}$. Additionally, $t[q_j,q_j+|s^{i_j}|-1]=s^{i_j}$ for $j\in\{1,\dots,r\}$. Note that  a sequence $(i_1,\dots,i_r)$ can have duplicates.

Informally, we want to construct $t$ from $S$ with possible overlapping. Let us denote the problem as $\texttt{TAO}(t,S)$.

\subsection{Connection with Real-World Problems}\label{sec:dna}
The considered problems have a strong relationship with the sequence assembly method for reconstructing a long DNA sequence from small fragments \cite{msdd2000}. Two types of the sequence assemble problem exist. The first one is de novo assembly. In this problem, we do not have the string $t$ and should construct it using all strings from the dictionary. This problem is NP-complete and typically is solved by heuristic algorithms. The SCS problem is one of the possible interpretations of this problem. 

The second type is Reference-guided genome assembly. This problem is similar to $\texttt{TAO}(t,S)$ problems but we should use a string from the dictionary only once. This differs from our problem where we can use strings from the dictionary $S$ several times. There are polynomial algorithms for the Reference-guided genome assembly problem that are presented in \cite{bo2017,ppds2004}.
 The current methods for DNA assembling belong to the ``next-generation'' \cite{vdd2009} or ``second-generation'' sequencing (NGS). They allow us to read many short substrings of DNA in a parallel way. Typically, a length of a DNA sequence (that is $t$ in our case) is about $10^8-10^9$ and the length of each piece is about $10^2-10^4$. 

The difference between our $\texttt{TAO}(t,S)$ and NGS problems is significant. At the same time, the NGS allows us to have several duplicates of a string in $S$. It is like relaxing the ``single usage of a string from $S$'' condition. It allow us to use a string from $S$ the fixed number of times. That is why our ideas can be used as a possible solution for the sequence assembly method in real-world examples.

An additional difference between the problems (in both cases de novo assembly and Reference-guided genome assembly) is the possibility of errors in the string $t$ for the case of DNA sequence assembling \cite{sb2007}. That is not allowed in our problems. That is why our algorithm should have improvement if it is used for DNA sequence assembling.

At the same time, our problems and algorithms are interesting as is because they solve fundamental problems and have applications in other areas like belonging a text to some natural language and others.
\subsection{Quantum Query Model}
We use the standard form of the quantum query model. 
Let $f:D\rightarrow \{0,1\},D\subseteq \{0,1\}^N$ be an $N$ variable function. An input for the function is $x=(x_1,\dots,x_N)\in D$ where $x_i\in\{0,1\}$ for $i\in\{1,\dots,N\}$.

We are given oracle access to the input $x$, i.e. it is implemented by a specific unitary transformation usually defined as $\ket{i}\ket{z}\ket{w}\rightarrow \ket{i}\ket{z+x_i\pmod{2}}\ket{w}$ where the $\ket{i}$ register indicates the index of the variable we are querying, $\ket{z}$ is the output register, and $\ket{w}$ is some auxiliary work-space. It can be interpreted as a sequence of control-not transformations such that we apply inversion operation (X-gate) to the second register that contains $\ket{z}$ in a case of the first register equals $i$ and the variable $x_i=1$. We interpret the oracle access transformation as $N$ such controlled transformations for each $i\in\{1,\dots,N\}$.   

An algorithm in the query model consists of alternating applications of arbitrary unitaries independent of the input and the query unitary, and a measurement in the end. The smallest number of queries for an algorithm that outputs $f(x)$ with a probability that is at least $\frac{2}{3}$ on all $x$ is called the quantum query complexity of the function $f$ and is denoted by $Q(f)$. We refer the readers to \cite{nc2010,a2017,aazksw2019part1,k2022lecturenotes} for more details on quantum computing. 

In this paper's quantum algorithms, we refer to the quantum query complexity as the quantum running time. We use modifications of Grover's search algorithm \cite{g96,bbht98} as quantum subroutines. For these subroutines, time complexity is more than query complexity for additional log factor \cite{ad2017,g2002}.

\section{Tools}\label{sec:tools}

Our algorithms use several data structures and algorithmic ideas like segment tree \cite{l2017guide}, suffix array \cite{mm90}, rolling hash \cite{kr87}, and prefix sum \cite{cormen2001}. Let us describe them in this section. 

\subsection{Rolling Hash for Strings Equality Checking}\label{sec:roll-hash}
The rolling hash was presented in \cite{kr87}.
For a string $u=(u_1,\dots,u_{|u|})$, we define a rolling hash function 
$h_p(u)=\left(\sum\limits_{i=1}^{|u|}u_i\cdot 2^{i-1}\right)\mbox{ mod }p,$
where $p$ is a prime. The presented implementation is for the binary alphabet but it can be easily extended for an arbitrary alphabet.

We can use the rolling hash and the fingerprinting method \cite{Fre79} for comparing two strings $u$ and $v$. The technique has many applications including quantum ones \cite{av2013,kk2017,kkk2022,av2008,aavz2016,an2009,af98,agky16,aakv2018,ag05}. Let us  randomly choose $p$ from the set of the first $r$ primes, such that $r\leq \frac{\max(|u|,|v|)}{\varepsilon}$ for some $\varepsilon>0$. Due to Chinese Remainder Theorem and \cite{Fre79}, if we have $h_p(u)=h_p(v)$ and $|u|=|v|$, then $u=v$ with error probability at most $\varepsilon$. If we compare $\delta$ different pairs of numbers, then we should choose an integer $p$ from the first $\frac{\delta \cdot \max(|u|,|v|)}{\varepsilon}$ primes for getting the error probability $\varepsilon$ for the whole algorithm. Due to Chebishev's theorem, the $r$-th prime number $p_r\approx r\ln r$. So, if our data type for integers is enough for storing $\frac{\delta \cdot \max(|u|,|v|)}{\varepsilon}\cdot (\ln(\delta) + \ln(\max(|u|,|v|))-\ln(\varepsilon))$, then it is enough for computing the rolling hash.


For a string $u$, we can compute a prefix rolling hash, that is
$h_p(u[1,i])$ for $i\in\{1,\dots,|u|\}$. It can be computed in $O(|u|)$ running time using the formula
\[h_p(u[1,i])=\left(h_p(u[1,i-1])+(2^{i-1}\mbox{ mod }p)\cdot u_i\right)\mbox{ mod }p\mbox{ and }h_p(u[1,0])=0 \mbox{ by definition}.\]
Assume that we store ${\cal K}_i=2^{i-1}$ mod $p$. We can compute all ${\cal K}_i$ in $O(|u|)$ running time using formula ${\cal K}_i=({\cal K}_{i-1}\cdot 2)$ mod $p$.

Similarly to prefix hashes we can define and compute suffix hashes $h_p(u[j,|u|])$ for each suffix $u[j,|u|]$.
\subsection{Segment Tree with Range Updates}\label{sec:segment-tree}

We consider a standard segment tree data structure \cite{l2017guide} for an array $b=(b_1,\dots, b_l)$ for some integer $l$. Assume that each element $b_i$ is a pair $(g_i,d_i)$, where $g_i$ is a target value, that is used in the segment tree, and $d_i$ is some additional value, that is used for another part of the algorithm.
The segment tree is a full binary tree such that each node corresponds to a segment of the array $b$. If a node $v$ corresponds to a segment $(b_{left},\dots, b_{right})$, then we store $\max(g_{left},\dots ,g_{right})$ in the node. A segment of a node is the union of segments that correspond to their two children. Leaves correspond to single elements of the array $b$. 

A segment tree for an array $b$ can be constructed in $O(l)$ running time. The data structure allows us to invoke the following requests in $O(\log l)$ running time.

\begin{itemize}
   \item 
    {\bf Range update.} It has four integer parameters $i, j, x, y$ ($1\leq i\leq j\leq l$). The procedure should assign $g_q\gets x$ and $d_q\gets y$ if $g_q<x$ for $i\leq q\leq j$. For this purpose, it goes down from the root and searches for nodes covered by the segment $[i,j]$. Then, the procedure updates these nodes by the new maximum. Let $\textsc{Update}(st,i,j,x,y)$ be a procedure that performs this operation for the segment tree $st$ in $O(\log l)$ running time. The detailed implementation is in Appendix A.
   \item 
  {\bf Push.} The procedure pushes all updates that were done by $\textsc{Update}$ procedure before from nodes down to leaves and update leaves with actual values. Note that updating leaves implies updating corresponding elements of the $b$ array. Let $\textsc{Push}(st)$ be a procedure that implements the operation for the segment tree $st$ in $O(l)$ running time. The detailed implementation is in Appendix A.
 \item 
{\bf Request.} For an integer $i$ ($1\leq i\leq l$), we check the leaf that corresponds to $b_i$ and return its value. Let $\textsc{Request}(st,i)$ be a function that returns $b_i$ from the segment tree $st$ in constant running time.
\end{itemize}

Let $\textsc{ConstructSegmentTree}(b)$ be a function that constructs and returns a segment tree for an array $b$ in $O(l)$ running time. 
 We refer the readers to \cite{l2017guide} for more details on the segment tree with range updates.

\subsection{Suffix Array}
A suffix array \cite{mm90} is an array $suf=(suf_1,\dots,suf_{l})$ for a string $u$ where $l=|u|$ is the length of the string. The suffix array is the lexicographical order for all suffixes of $u$. Formally, $u[suf_i,l]<u[suf_{i+1},l]$ for any $i\in\{1,\dots,l-1\}$.
Let $\textsc{ConstructSuffixArray}(u)$ be a procedure that constructs the suffix array for the string $u$. The running time of the procedure is as follows:
\begin{lemma}[\cite{llh2018}]\label{lm:suf-arr}
A suffix array for a string $u$ can be constructed in $O(|u|)$ running time.
\end{lemma}

\section{Shortest Common Superstring Problem}\label{sec:csc}

We discuss our algorithm for the SCS problem in this section.
Assume that we have a pair $i$ and $j$ such that $s^{i}$ is a substring of $s^{j}$. In that case, if a superstring $t$ contains the string $s^{j}$ as a substring, then $t$ contains the string $s^{i}$ too. Therefore, we can exclude $s^{i}$ from the sequence $S$, and it does not affect the solution. Excluding such strings is the first step of the algorithm. 
In the rest part of the section, we assume that no string $s^i$ is a substring of any string $s^j$ for $i,j\in\{1,\dots,n\}$.

Secondly, let us reformulate the problem in a graph form. Let us construct a complete directed weighted graph $G=(V,E)$ by the sequence $S$. 
A node $v^i$ corresponds to the string $s^i$ for $i\in\{1,\dots,n\}$. The set of nodes (vertexes) is $V=(v^1,\dots,v^n)$. The weight of an edge between two nodes $v^i$ and $v^j$ is the length of the maximal overlap for $s^i$ and $s^j$. Formally, 
\[w(i,j)=\max\limits_{1\leq r\leq \min\{|s^i|,|s^j|\}}\{r: s^i[|s^i|-r+1,|s^i|]=s^j[1,r]\}.\]
We can see that any path that visits all nodes exactly once represents a superstring. Note that no string is a substring of another one. That is why, we cannot exclude any node from the path that corresponds to a superstring.
Let $P=(v^{i_1},\dots,v^{i_\ell})$ be a path. Let the weight of the path $P$ be $w(P)=w(v^{i_1},v^{i_2})+\dots+w(v^{i_{\ell-1}},v^{i_{\ell}})$ that is the sum of weights of all edges from $P$; let $|P|=\ell$.
The path that visits all nodes exactly once and has maximal weight represents the shortest superstring.
We formulate the above discussion as the following lemma and present its proof in Appendix D for completeness:

\begin{lemma}\label{lm:scs-graph-connection}
The path $P$ that visits all nodes of $G$ exactly once and has the maximal possible weight corresponds to the shortest common superstring $t$ for the sequence $S$.
(See Appendix D for the proof).
\end{lemma}

In fact, the mentioned problem on the graph $G$ is the Travelling Salesman Problem on a complete graph. The quantum algorithm for the TSP was developed in \cite{abikpv2019}. At the same time, the algorithm in \cite{abikpv2019} skips some details. Here we present the algorithm with all details to have detailed complexity (in this paper we are interested even in log factors) and for completeness of presentation. At the same time, before TSP we have several subproblems that should be solved for SCS and have not quantum algorithms yet. These subproblems are removing substrings and constructing the graph. They are discussed in the remaining part of the section.

Let us present three procedures:
\begin{itemize}
    \item $\textsc{RemovingDublicatesAndSubstrings}(S)$ is the first step of the algorithm that removes any duplicates from $S$ and strings that are substrings of any other strings from $S$. The implementation of the procedure is presented in Algorithm \ref{alg:duble}. The algorithm is following. Firstly, we sort all strings of $S$ by the length in ascending order. Secondly, for each string $s^i$ we check whether it is a substring of $\tilde{s}^i$. Here \[\tilde{s}^i=s^{i+1} \$ \dots \$ s^{n},\] that is the concatenation of all strings from $S$ with indexes bigger than $i$ using $\$$ symbol as a separator, and $\$$ is a symbol that cannot be in any string. If $s^i$ is a substring of $\tilde{s}^i$, then there is $j>i$ such that $s^i$ is a  substring of $s^j$ or $s^i=s^j$ because all strings are separated by ``non-alphabetical'' symbol. 
    
    The implementation uses $IsSubstring(s^i,\tilde{s}^i)$ subroutine that returns $True$ if $s^i$ is a substring of $\tilde{s}^i$ and $False$ otherwise. We use quantum algorithm for the strings matching problem \cite{rv2003} with  $O^*\left(\sqrt{|s^i|}+\sqrt{|\tilde{s}^i|}\right)$ running time. So, if the procedure returns true, we can exclude the string from the final sequence $S$.
    
    For access to $\tilde{s}^i$, we do not need to concatenate these strings. It is enough to implement $\textsc{GetSymbol}(i,j)$ function that returns $j$-th symbol of $\tilde{s}^i$. The index of the string $i\in\{1,\dots,n\}$, and the index of the symbol $j\in\{1,\dots,|s^{i+1}|+\dots+|s^{n}|+i-2\}$. The implementation of the function is based on Binary search and has $O(\log n)$ running time, it is presented in Algorithm \ref{alg:get-symbol}. The subroutine requires precalculated array $start$ such that $start[i] = |s^1|+\dots+|s^{i-1}|+1$ for $i\in\{1,\dots,n\}$, and $start[1] = 1$. It can be computed in $O(n)$ running time using formula $start[i] = start[i-1]+|s^{i-1}|$ for $i\in\{2,\dots,n\}$.  Complexity of the $\textsc{RemovingDublicatesAndSubstrings}(S)$ procedure is discussed in Lemma \ref{lm:remove-dublicated}.
    
    \item $\textsc{ConstructTheGraph}(S)$ constructs the graph $G=(V,E)$ by $S$. The main idea is the following one. We randomly choose a prime $p$ among the first $20nL$ primes. Then, we compute the rolling hash function with respect to $p$ for each prefix and suffix of $s^i$, where $i\in\{1,\dots,n\}$. Assume that we have $\textsc{ComputePrefixAndSuffixHashes}(s^i,p)$ subroutine for computing the prefix and suffix hashes and storing them in an array. After that, we can take the result of the rolling hash for any prefix or suffix of $s^i$ in constant running time. 
    
    For each pair of strings $s^i$ and $s^j$ we define a search function $sp_{i,j}:\{0,\dots,min(|s^i|,|s^j|)\}\to\{0,1\}$ such that $sp_{i,j}(r)=1$ iff $h_p(s^i[r+1,|s^i|])=h_p(s^j[1,r])$ that means $s^i[r+1,|s^i|]=s^j[1,r]$ with high probability. We define $sp_{i,j}(0)=1$. In fact, $w(v^i,v^j)$ is the maximal $1$-result argument of $sp_{i,j}$. We can find it using First One Search algorithm \cite{dhhm2006,k2014,ll2015,ll2016,kkmsy2022} in $O(\sqrt{min(|s^i|,|s^j|)})$ running time.
    
    The implementation of the subroutine is presented in Algorithm \ref{alg:constr}. Here we assume that we have $\textsc{FirstOneSearch}(sp_{i,j})$ subroutine. Complexity of the procedure is discussed in Lemma \ref{lm:construst-graph}
    \item $\textsc{ConstructSuperstringByPath}(P)$ constructs the target superstring by a path $P$ in the graph $G=(V,E)$. Implementation of the procedure is presented in Algorithm \ref{alg:getpath}.
\end{itemize}
\begin{algorithm}
\caption{Implementation of $\textsc{RemovingDublicatesAndSubstrings}(S)$ for $S=(s^1,\dots,s^n)$.}\label{alg:duble}
\begin{algorithmic}
\State $setOfDeletingIndexes\gets\{\}$\
\State $\textsc{SortByLength}(S)$
\For{$i\in\{1,\dots,n\}$}
\If{$\textsc{IsSubstring}(s^i,\tilde{s}^i)=True$}\Comment{$s^i$ is a substring of $\tilde{s}^i$}
\State $setOfDeletingIndexes\gets setOfDeletingIndexes\cup\{i\}$
\EndIf
\EndFor
\For{$i\in setOfDeletingIndexes$}\Comment{The running time of the for-loop is $O(n)$}
\State Remove $s^i$ from $S$
\EndFor
\State $n\gets n - |setOfDeletingIndexes|$\Comment{We update $n$ by the actual value that is the size of $S$.}

\end{algorithmic}
\end{algorithm}

\begin{algorithm}
\caption{Implementation of $\textsc{GetSymbol}(i,j)$ for $i\in\{1,\dots,n\},j\in\{1,\dots,|s^{i+1}|+\dots+|s^n|+i-2\}$.}\label{alg:get-symbol}
\begin{algorithmic}
\State $j\gets j-1+start[i+1] + i$\Comment{We update all symbol indexes by indexes in the concatenation of all strings from $S$ via $\$$-separator.}
\State $left \gets i+1$, $right\gets n$
\State $symbol \gets NULL$\Comment{The symbol is unknown at the moment}
\While{$left\leq right$}
\State $mid\gets \lfloor (left+right)/2 \rfloor$
\State $jm\gets start[mid]+(mid-1)$\Comment{The index of the first symbol of $s^{mid}$}
\If{$j\geq jm$ and $j<jm + |s^{mid}|$}\Comment{$j$ is inside $s^{mid}$ }
\State $j'\gets j - jm +1$
\State $symbol\gets s^{mid}_{j'}$
\State stop the while loop.
\EndIf
\If{$j= jm + |s^{mid}|$}\Comment{the $j$-th symbol is separator}
\State $symbol\gets \$$
\State stop the while loop.
\EndIf
\If{$j < jm$}
\State $right\gets mid-1$
\Else
\State $left\gets mid+1$
\EndIf
\EndWhile
\State \Return{$symbol$}
\end{algorithmic}
\end{algorithm}

\begin{algorithm}
\caption{Implementation of $\textsc{ConstructTheGraph}(S)$ for $S=(s^1,\dots,s^n)$.}\label{alg:constr}
\begin{algorithmic}
\State $V=(v^1,\dots,v^n)$
\State $p\in_R\{p_1,\dots,p_{20nL}\}$\Comment{We randomly choose a prime $p$ among the first $20nL$ primes}
\For{$i\in\{1,\dots,n\}$}
 \State $\textsc{ComputePrefixAndSuffixHashes}(s^i,p)$
\EndFor
\For{$i\in\{1,\dots,n\}$}
\For{$j\in\{1,\dots,n\}$}
\If{$i\neq j$}
\State $maxOverlap\gets \textsc{FirstOneSearch}(sp_{i,j})$
\State $E\gets E\cup \{(v^i,v^j)\}$
\State $w(v^i,v^j)\gets maxOverlap$
\EndIf
\EndFor
\EndFor
\State \Return{$(V,E)$}
\end{algorithmic}
\end{algorithm}

\begin{algorithm}
\caption{Implementation of $\textsc{ConstructSuperstringByPath}(P)$ for $P=(v^{i_1},\dots,v^{i_\ell})$.}\label{alg:getpath}
\begin{algorithmic}
\State $t=s^{i_1}$
\For{$j\in\{2,\dots,\ell\}$}
\State $t\gets t\circ s^{i_j}[w(v^{i_{j-1}},v^{i_{j-1}})+1,|s^{i_j}|]$\Comment{Here $\circ$ is the concatenation operation.}
\EndFor
\State \Return{$t$}
\end{algorithmic}
\end{algorithm}

\begin{lemma}\label{lm:remove-dublicated}
The $\textsc{RemovingDublicatesAndSubstrings}(S)$ procedure removes duplicates and substrings in $O(n\sqrt{L}\log^2 L\log^2 n)$ running time with the error probability at most $0.1$.
\end{lemma}
\Beginproof
Firstly, we sort all strings by the length that takes $O(n\log n)$ running time. Then, we invoke $\textsc{IsSubstring}(s^i,\tilde{s}^i)$ procedure $n$ times. Due to \cite{rv2003} and complexity of the Binary search algorithm for $\textsc{GetSymbol}(i,j)$, each invocation has the following running time \[O\left(\sqrt{|\tilde{s}^i|}\log\sqrt{\frac{|\tilde{s}^i|}{|s^i|}}\log|s^i|\log n+\sqrt{|s^i|}\log^2|s^i|\right).\]

Note that $|\tilde{s}^i|=O(|s^{i+1}|+\dots+|s^{n}|)\leq O(|s^{1}|+\dots+|s^{i-1}|+|s^{i+1}|+\dots+|s^{n}|)=O(L-|s^i|)$. Therefore, according to  the Cauchy--Bunyakovsky--Schwarz inequality we have the following complexity:
\[= O\left( \sqrt{L-|s^i|}\log L \log |s^i|\log n+\sqrt{|s^i|}\log^2|s^i|\right)= O\left( (\sqrt{L-|s^i|}+\sqrt{|s^i|})\log L \log |s^i|\log n\right)\]\[=  O\left( \sqrt{2\cdot(L-|s^i|+|s^i|)}\log L \log |s^i|\log n\right)=O\left( \sqrt{L}\log L \log |s^i|\log n\right)\]

The total complexity is $O\left(n\log n+ \sqrt{L}\log L \log |s^i|\log n\right)=O\left( \sqrt{L}\log L \log |s^i|\log n\right)$.

Each invocation of $\textsc{IsSubstring}(s^i,\tilde{s}^i)$ has constant error probability and it can be accumulated during $n$ invocations. That is why we repeat each invocation $O(\log n)$ times for obtaining constant total error probability, for example at most $0.1$ total error probability can be achieved.

So, the total complexity of the procedure is
\[O\left(\sum_{i=1}^n \sqrt{L}\log L \log |s^i|\log^2 n\right)=O\left(n\sqrt{L}\log L \log\left(\max_{i=\{1,\dots,n\}}|s^i|\right)\log^2 n\right)=O(n\sqrt{L}\log^2 L\log^2 n).\]
\Endproof

\begin{lemma}\label{lm:construst-graph}
The $\textsc{ConstructTheGraph}(S)$ procedure constructs the graph in $O(L+n^{1.5}\sqrt{L})$ running time with the error probability at most $0.1$.
\end{lemma}
\Beginproof
Computing prefix and suffix hashes for a string $s^i$ have $O(|s^i|)$ running time. So, computing them for all strings has $O(|s^1|+\dots+|s^n|)=O(L)$ running time. 

Complexity of computing overlaps for a fixed $s^i$ is at most $O(n\sqrt{|s^i|})$. Therefore, the total complexity of computing all overlaps for all strings is 
\[O(n\sqrt{|s^1|}+\dots+n\sqrt{|s^n|})= O(n(\sqrt{|s^1|}+\dots+\sqrt{|s^n|}))=O(n\sqrt{n(|s^1|+\dots+|s^n|)})=O(n^{1.5}\sqrt{L}).\]

So the total complexity is $O(L+n^{1.5}\sqrt{L})$.

Each $\textsc{FirstOneSearch}$ invocation has an error probability. Therefore, the total error probability can be close to $1$. At the same time, the algorithm is a sequence of  First One Search algorithms that can be converted to an algorithm with constant error probability at most $0.05$ without affecting running time using \cite{k2014,lmrss2011} technique.

For a fixed $s^i$, the number of hash comparisons is at most $O(n|s^i|)$ that is at most $n$ different suffixes for each prefix of $s^i$. Therefore, the total number of different pairs of numbers that are compared is 
$O(n|s^1|+\dots+n|s^n|)= O(nL).$
Hence, if we randomly choose a prime $p$ among the first  $20nL$, then we can archive the error probability at most $0.05$ of all computations. 
So, the total error probability is at most $0.1$. 
\Endproof

Let us discuss the main part of the algorithm.
We consider a function $L:2^V\times V\times V\to \mathbb{R}$ where $2^V$ is the set of all subsets of $V$. The function $L$ is such that $L(Y,v,u)$ is the maximum of all weights of paths that visit all nodes from $Y$ exactly once, start from the node $v$, and finish in the node $u$. If there is no such path, then we assume that $L(Y,v,u)=-\infty$.

Let the function $F:2^V\times V\times V\to V^*$ be such that $F(Y,v,u)$ is the path that visits all nodes of $Y$ exactly once, starts from the node $v$, finishes in the node $u$ and has the maximal weight. In other words, for $P=F(Y,u,v)$ we have $w(P)=L(Y,u,v)$.
We assume, that $L(\{v\},v,v)=0$  and $F(\{v\},v,v)=(v)$ for any $v\in V$ by definition.

Let us discuss properties of the function.

\begin{property} Suppose $Y\subseteq V, v,u\in Y$, an integer $k< |Y|$. The function $L$ is such that
\[L(Y,v,u)=\max\limits_{Y'\in\{Y':Y'\subset Y,|Y'|=k, v\in Y', u\not\in Y'\}}\left\{\max\limits_{y\in Y'}\left\{L(Y',v,y)+L((Y\backslash Y') \cup \{y\},y,u)\right\}\right\}\]
and $F(Y,u,v)$ is the path that is concatenation of corresponding paths.
\end{property}
\Beginproof
Let us fix a set $Y'$ such that $Y'\subset Y,$ $|Y'|=k$.
Let $P^1(Y')=F(Y',v,y_{max}(Y'))$ and $P^2(Y')=F((Y\backslash Y') \cup \{y\},y_{max}(Y'),u)$, where $y_{max}(Y')$ is the target argument for the inner maximum.

The path $P(Y')=P^1(Y')\circ P^2(Y')$ belongs to $Y$, starts from $v$ and finishes in $u$, where $\circ$ means concatenation of paths excluding the duplication of common node $y_{max}(Y')$. 

Let us consider all paths such that they visit all elements from $Y'$, then other elements of $Y$. In other words, paths $T=(v^{i_1},\dots,v^{i_\ell})$ such that $\{v^{i_1},\dots,v^{i_k}\}=Y'$, and $\{v^{i_{k+1}},\dots,v^{i_\ell}\}=Y\backslash Y'$. So, $v^{i_k}\in Y'$, and $\{v^{i_{k}},\dots,v^{i_\ell}\}=Y\backslash Y'\cup\{v^{i_k}\}$. Therefore, due to selecting $y_{max}(Y')$ as a target element for maximum, we can be sure that $w(P(Y'))\geq w(T)$.

Let $P=P(Y'_{max})$ such that we reach the outer maximum on $Y'_{max}$. It belongs to $Y$, starts from $v$, and finishes in $u$. Therefore, $w(P)\leq L(Y,v,u)$. 

Assume that there is a path $T=(u^{i_1},\dots,u^{i_\ell})$ such that $w(T)=L(Y,v,u)$ and $w(T)>w(P)$. Let us select $Y''=\{u^{i_1},\dots,u^{i_k}\}$. So, it is such that $Y''\subset Y$ and $|Y''|=k$. Due to the above discussion $T'=P(Y'')$. Therefore, $w(P(Y''))>w(P(Y'_{max}))$ contradicts the definition of $P(Y'_{max})$ as a path where we reach the outer maximum.
\Endproof

As a corollary, we obtain the following result. Note that each pair of edges is connected.
\begin{corollary}\label{cor:one-edge}
Suppose $Y\subset V,$ $v,u\in Y$. The function $L$ is such that
\[L(Y,v,u)=\max\limits_{y\in Y\backslash\{u\}}\left(L( Y\backslash\{u\},v,y)+w(y,u)\right).\]
and $F(Y,u,v)$ is the corresponding path.
\end{corollary}

Using this idea, we construct the following algorithm.

{\bf Step 1}. Let $\alpha=0.055$. We classically compute $L(S,v,u)$ and $F(S,v,u)$ for all $Y\subset V$ such that $|Y|\leq (1-\alpha)\frac{n}{4}$ and $v,u\in Y$ 

{\bf Step 2}. Let $V_4\subset V$ be such that $|V_4|=\frac{n}{4}$. Then, we have 

\[L(V_4,u,v)=\max\limits_{V_{\alpha}\in\{V_{\alpha}:V_{\alpha}\subset V_4,|V_{\alpha}|=(1-\alpha)\frac{n}{4}\},y\in V_{\alpha}}\left(L(V_{\alpha},v,y)+L((V_4\backslash V_{\alpha}) \cup \{y\},y,u)\right).\]

Let $V_2\subset V$ be such that $|V_2|=\frac{n}{2}$. Then, we have

\[L(V_2,u,v)=\max\limits_{V_{4}\in\{V_{4}:V_4\subset V_2,|V_4|=\frac{n}{4}\},y\in V_4}\left(L(V_4,v,y)+L((V_2\backslash V_4) \cup \{y\},y,u)\right).\]

Finally,
\[L(V,u,v)=\max\limits_{V_{2}\in\{V_{2}:V_2\subset V,|V_2|=\frac{n}{2}\},y\in V_2}\left(L(V_2,v,y)+L((V\backslash V_2) \cup \{y\},y,u)\right).\]

We can compute $L(V,u,v)$ and corresponding $F(V,u,v)$ using three nested procedures for maximum finding. As such procedure, we use D\"urr-H\o yer \cite{dh96,dhhm2006} quantum minimum finding algorithm. The maximal weight of paths $MaxWeight$ and the corresponding path can be computed as a maximum of $L(V,u,v)$ among all $u,v\in V$ as presented in the next statement.

\[MaxWeight=\max\limits_{u,v\in V} L(V,v,u).\]

Let us discuss the implementation of Step 1. It is presented as a recursive function $\textsc{GetL}(Y,v,u)$ for $Y\subset V,u,v\in V$ with cashing that is Dynamic Programming approach in fact. The function is based on Corollary \ref{cor:one-edge}.

\begin{algorithm}[ht]
\caption{$\textsc{GetL}(Y,v,u)$.}
\begin{algorithmic}
\If{$v=u$ and $Y=\{v\}$}\Comment{Initialization}
\State $L(\{v\},v,v)\gets 0$
\State $F(\{v\},v,v)\gets (v)$
\EndIf
\If {$L(Y,v,u)$ is not computed}
\State $weight\gets -\infty$
\State $path\gets ()$
\For{$y\in Y\backslash\{u,v\}$}
\If{$\textsc{GetL}( Y\backslash\{u\},v,y)+w(y,u)>weight$}
\State $weight\gets L( Y\backslash\{u\},v,y)+w(y,u)$
\State $path\gets F( Y\backslash\{u\},v,y)\cup u$
\EndIf
\EndFor
\State $L(Y,v,u)\gets weight$
\State $F(Y,v,u)\gets path$
\EndIf
\State \Return{$L(Y,v,u)$}
\end{algorithmic}
\end{algorithm}

\begin{algorithm}
\caption{$\textsc{Step1}$.}
\begin{algorithmic}
\For{$Y \in 2^V$ such that $|Y|\leq (1-\alpha)\frac{n}{4}$}
\For{$v\in Y$}
\For{$u\in Y$}
\State $\textsc{GetL}(Y,v,u)$\Comment{We are computing $L(Y,v,u)$ and $F(Y,v,u)$ for Step 2.}
\EndFor
\EndFor
\EndFor
\end{algorithmic}
\end{algorithm}

Let $\textsc{QMax}((x_1,\dots,x_N))$ be the implementation of the quantum maximum finding algorithm \cite{dh96,dhhm2006} for a sequence $(x_1,\dots,x_N)$.
The most nested quantum maximum finding algorithm for some $V_4\subset V, |V_4|=\frac{n}{4}$ and $u,v\in V_4$ is \[\textsc{QMax}((L(V_{\alpha},v,y)+L((V_4\backslash V_{\alpha}) \cup \{y\},y,u):V_{\alpha}\subset V_4,|V_{\alpha}|=(1-\alpha)\frac{n}{4},y\in V_{\alpha})).\]

The second quantum maximum finding algorithm for some $V_2\subset V, |V_2|=\frac{n}{2}$ and $u,v\in V_2$ is 
\[\textsc{QMax}((L(V_{4},v,y)+L((V_2\backslash V_{4}) \cup \{y\},y,u):V_4\subset V_2,|V_4|=n/4,y\in V_4)).\]
Note that $|V_4|=n/4$ and $|V_2\backslash V_{4}|=n/4$. We use the invocation of $\textsc{QMax}$ (the most nested quantum maximum finding algorithm) instead of $L(V_{4},v,y)$  and $L(V_2\backslash V_{4},y,u)$.

The third quantum maximum finding algorithm for some  $u,v\in V$ is 

\[\textsc{QMax}((L(V_2,v,y)+L((V\backslash V_2) \cup \{y\},y,u):V_2\subset V,|V_2|=n/2,y\in V_2))\]

Note that $|V_2|=n/2$ and $|V\backslash V_2|=n/2$. We use the invocation of $\textsc{QMax}$ (the second quantum maximum finding algorithm) instead of $L(V_2,v,y)$ and $L((V\backslash V_2) \cup \{y\},y,u)$.

The fourth quantum maximum finding algorithm among all  $u,v\in V$ is 

\[\textsc{QMax}(L(V,v,u):v,u\in V)\]

The procedure $\textsc{QMax}$ returns not only the maximal value but the index of the target element. Therefore, by the ``index'' we can obtain the target paths using the $F$ function. So, the resulting path is
$P=P^1\circ P^2$, where $P^1$ is the result path for $L(V_2,v,y)$ and $P^2$ is the result path for $L((V\backslash V_2) \cup \{y\},y,u)$.

$P^1=P^{1,1}\circ P^{1,2}$, where $P^{1,1}$ is the result path for $L(V_{4},v,y)$ and $P^{1,2}$ is the result path for $L((V_2\backslash V_{4}) \cup \{y\},y,u)$. In the same way, we can construct $P^2=P^{2,1}\circ P^{2,2}$.

$P^{1,1}=P^{1,1,1}\circ P^{1,1,2}$, where $P^{1,1,1}$ is the result path for $L(V_{\alpha},v,y)$ and $P^{1,1,2}$ is the result path for $L((V_4\backslash V_{\alpha}) \cup \{y\},y,u)$. Note, that these values were precomputed classically in Step 1, and were stored in $F(V_{\alpha},v,y)$ and $F((V_4\backslash V_{\alpha}) \cup \{y\},y,u)$ respectively.

In the same way, we can construct
\[P^{1,2}=P^{1,2,1}\circ P^{1,2,2},\quad\quad
P^{2,1}=P^{2,1,1}\circ P^{2,1,2},\quad\quad
P^{2,2}=P^{2,2,1}\circ P^{2,2,2}.\]

The final path is
\[P=P^1\circ P^2=(P^{1,1}\circ P^{1,2})\circ(P^{2,1}\circ P^{2,2})=\]
\[\Big((P^{1,1,1}\circ P^{1,1,2})\circ (P^{1,2,1}\circ P^{1,2,2})\Big)\circ\Big((P^{2,1,1}\circ P^{2,1,2})\circ (P^{2,2,1}\circ P^{2,2,2})\Big)\]

Note the Durr-Hoyer algorithm $\textsc{QMax}$ has an error probability at most $0.1$ and is based on the Grover search algorithm. So, because of several nested $\textsc{QMax}$ procedures, we should use the bounded-error input version of the Grover search algorithm that was discussed in \cite{hmw2003,abikkpssv2020,abikkpssv2023}.

Let us present the final algorithm as Algorithm \ref{alg:main}. The complexity of the algorithm is presented in Theorem \ref{th:scs-compl}.
\begin{algorithm}
\caption{Algorithm for $\texttt{SCS}(S)$.}\label{alg:main}
\begin{algorithmic}
\State $\textsc{RemovingDublicatesAndSubstrings}(S)$
\State $(V,E)\gets\textsc{ConstructTheGraph}(S)$
\State $\textsc{Step1}()$
\State $weight, path \gets \textsc{QMax}(L(V,v,u):v,u\in V)$
\State $t\gets \textsc{ConstructSuperstringByPath}(path)$
\State \Return{t}
\end{algorithmic}
\end{algorithm}

\begin{theorem}
 Algorithm \ref{alg:main} solves $\texttt{SCS}(S)$ with \[O\left(n^{3}1.728^n+L+n^{1.5}\sqrt{L}+n\sqrt{L}\log^2L\log^2n\right)\] running time and error probability at most $1/3$.
\end{theorem}\label{th:scs-compl}
\Beginproof
The correctness of the algorithm follows from the above discussion. Let us present an analysis of running time.
The complexity of removing all duplicates and substrings in $S$ by the procedure $\textsc{RemovingDublicatesAndSubstrings}(S)$ is $O(n\sqrt{L}\log^2 L\log^2 n)$ and the error probability is at most $0.1$ due to Lemma \ref{lm:remove-dublicated}.
The complexity of constructing the graph is $O(L+n^{1.5}\sqrt{L})$ and the error probability is at most $0.1$ due to Lemma \ref{lm:construst-graph}.

We continue with complexity of Step 1 (Classical preprocessing).  Here we check all subsets of size at most $(1-\alpha) \frac{n}{4}$, starting and ending nodes, and neighbor nodes. 
The running time is 

\[\sum_{i=1}^{(1-\alpha) \frac{n}{4}}O\left( \binom{n}{i}n^3\right)=O(n^3 1.728^n).\]

Complexity of Step 2 (Quantum part) is the complexity of four nested Durr-Hoyer maximum finding algorithms. Due to  \cite{dh96,g96,dhhm2006}, the complexity for the most nested $\textsc{QMax}$ is
\[Q_1=O\left(\sqrt{\binom{n/4}{\alpha n/4}}\cdot \sqrt{n}\right)\] because the searching space size is $\binom{n/4}{\alpha n/4}\cdot n$ and the running time of extracting the subset form its number is $O(n)$. These two operations are invoked sequentially.
The complexity for the second $\textsc{QMax}$ is
\[Q_2=\left(\sqrt{\binom{n/2}{n/4}}\cdot \sqrt{n}\cdot Q_1\right)=O\left(\sqrt{\binom{n/2}{n/4}\binom{n/4}{\alpha n/4}}\cdot n\right)\] because of the similar reasons.
The complexity for the third $\textsc{QMax}$ is
\[Q_3 = O\left(\sqrt{\binom{n}{n/2}}\cdot \sqrt{n}\cdot Q_2\right)=O\left(\sqrt{\binom{n}{n/2}\binom{n/2}{n/4}\binom{n/4}{\alpha n/4}}\cdot n^{1.5}\right).\]
The complexity for the fourth (that is the final) $\textsc{QMax}$ is
\[O\left(\sqrt{n^2}\cdot Q_3\right)=O\left(\sqrt{\binom{n}{n/2}\binom{n/2}{n/4}\binom{n/4}{\alpha n/4}}\cdot n^{2.5}\right)\] because the searching space size is $n^2$.

So, the total complexity of Step 2 is 
\[O\left(\sqrt{\binom{n}{n/2}\binom{n/2}{n/4}\binom{n/4}{\alpha n/4}}\cdot n^{2.5}\right)=O(n^{2.5}1.728^n).\]

The complexity of $\textsc{ConstructSuperstringByPath}$ is $O(L)$.

We invoke $\textsc{RemovingDublicatesAndSubstrings}$, $\textsc{ConstructTheGraph}$, Step 1, Step 2 and $\textsc{ConstructSuperstringByPath}$  sequentially. Therefore, the total complexity is the sum of complexities for these steps. So, the total complexity is 
\[O\left(n\sqrt{L}\log^2L\log^2n+L+n^{1.5}\sqrt{L}+n^{3}1.728^n+n^{2.5}1.728^n + L\right)\]\[=O\left(n^{3}1.728^n+L+n^{1.5}\sqrt{L}+n\sqrt{L}\log^2L\log^2n\right).\]

Step 2, $\textsc{RemovingDublicatesAndSubstrings}$ and $\textsc{ConstructTheGraph}$ have error probability. Each of them has at most constant error probability. Using repetition, we can achieve at most $0.1$ error probability for each of the procedures and at most $0.3$ for the whole algorithm.
\Endproof

\subsection{New Algorithm for Constructing Graph}
In this Section, we discuss an alternative implementation of $\textsc{ConstructTheGraph}$ procedure that is $\textsc{ConstructTheGraph2}(S)$ that constructs the graph $G=(V,E)$ by $S$. For the algorithm we need a quantum procedure $\textsc{AllOnesSearch}(i,\mathcal{I},r)$ that
\begin{itemize}
    \item accepts an index of a string $i\in\{1,\dots,n\}$, a set of indexes of strings $\mathcal{I}\subset\{1,\dots,n\}$, and an index of a symbol from a string $r\in\{1,\dots,|s^i|\}$.
    \item returns a set of indexes $\mathcal{I}'\subset \mathcal{I}$ such that for any $j\in\mathcal{I}'$ we have $s^{i}[n-r+1]= s^j[r]$.
\end{itemize}
The function is based on Grover's search algorithm \cite{g96,bbht98} and has $O\left(\sqrt{|\mathcal{I}|\cdot |\mathcal{I}'|}\right)$ complexity. The procedure and analysis are presented in \cite{k2014}.
 
The main idea of the algorithm is the following. 

\begin{itemize}
    \item \textbf{Step 1}. Initially, we assign $w(v^i,v^j)\gets 0$ for each $i,j\in\{1,\dots,n\}$.
    \item \textbf{Step 2}. We consider all strings $s^i$, for $i\in\{1,\dots, n\}$. For each string $s^i$ we start from $\mathcal{I} \gets\{1,\dots,n\} \backslash \{i\}$, and $r\gets 1$. We do Step 3. until $\mathcal{I}= \emptyset$.
    \item \textbf{Step 3}. For 
$i\in\{1,\dots,n\}$,  $\mathcal{I}\subset\{1,\dots,n\}$, and  $r\in\{1,\dots,|s^i|\}$  
we invoke $\mathcal{I}'\gets \textsc{AllOnesSearch}(i,\mathcal{I},r)$. Then, we update $w(v^i,v^j)\gets w(v^i,v^j)+1$ for each $j\in \mathcal{I}'$. After that, if $\mathcal{I}'\neq \emptyset$ we repeat Step 3 for $\mathcal{I}\gets \mathcal{I}'$, and $r\gets r+1$. If $\mathcal{I}'=\emptyset$ then we stop the process for the current $s^i$.
\end{itemize}

The implementation of the procedure is presented in Algorithm \ref{alg:newcon}. The complexity of the procedure is discussed in Lemma \ref{lm:newcon}.

\begin{algorithm}
\caption{Implementation of $\textsc{ConstructTheGraph2}(S)$ for $S=(s^1,\dots,s^n)$.}\label{alg:newcon}
\begin{algorithmic}
\State $V=(v^1,\dots,v^n)$
\For{$i\in\{1,\dots,n\}$}
\For{$j\in\{1,\dots,n\}$}
\If{$i\neq j$}
 \State $w(v^i,v^j)\gets 0$
\EndIf
\EndFor
\EndFor
\For{$i\in\{1,\dots,n\}$}
\State $\mathcal{I} \gets\{1,\dots,n\} \backslash \{i\}$
\State $r\gets 1$
\State $\mathcal{I}\gets \textsc{AllOnesSearch}(i,\mathcal{I},r)$
\While{$\mathcal{I}'\neq \emptyset$}
\For{$j\in \mathcal{I}$}
\State $w(v^i,v^j)\gets w(v^i,v^j)+1$ 
\EndFor
\State $\mathcal{I}\gets \textsc{AllOnesSearch}(i,\mathcal{I},r)$
\State $r\gets r+1$
\EndWhile
\EndFor
\State \Return{$(V,E)$}
\end{algorithmic}
\end{algorithm}

The procedure has good complexity in the case of $s^i$ are random.
\begin{lemma}\label{lm:newcon}
Assume that the alphabet $\Sigma$ has order $|\Sigma| = C$ and all strings $s^i$ are random. The $\textsc{ConstructTheGraph2}(S)$ procedure constructs the graph in $O(n^{2}\cdot\frac{C^{1/2}}{C-1})$ running time on average with the error probability at most $0.1$.
\end{lemma}

\Beginproof
We use the All Ones Search Problem to search all $k\in \mathcal{I}$ such that $s^{i}[n-r+1]= s^k[r]$ and the complexity of it is $O\left(\sqrt{|\mathcal{I}|\cdot |\mathcal{I}'|}\right)$, where $|\mathcal{I}|,|\mathcal{I}'|\leq n-1$. In this case we have $m=|\mathcal{I}|$ with. Since the alphabet $\Sigma$ has order $C$ and all strings $s^i$ are random, the output of $\textsc{AllOnesSearch}(i,\mathcal{I},r)$ has order $\frac{1}{C}\cdot |\mathcal{I}|$ on average. Therefore, the total complexity of constructing the graph is 
\[ n \cdot \sum_{k=1}^{L} O\left(\sqrt{\frac{1}{C}\cdot\frac{n-1}{C^{k-1}} \cdot\frac{n-1}{C^{k-1}}  }\right) = O\left(n^2\cdot \frac{C^{1/2}}{C-1}\right)\]
\Endproof
The complexity of whole algorithm in that case is presented in the next Corollary
\begin{corollary}
    Assume that the alphabet $\Sigma$ has order $|\Sigma| = C$ and all strings $s^i$ are random. Algorithm \ref{alg:main} with this assumption solves $\texttt{SCS}(S)$ with \[O\left(n^{3}1.728^n+n^{2}\cdot\frac{C^{1/2}}{C-1}+n\sqrt{L}\log^2L\log^2n + L\right)\] running time in avarage and error probability at most $1/3$, where $C$. In the case of $C=const$, the running time is 
    \[O\left(n^{3}1.728^n+n\sqrt{L}\log^2L\log^2n + L\right).\] 
\end{corollary}
\Beginproof
Due to Lemma \ref{lm:newcon}, the complexity of  graph constructing is $O(n^{2}\cdot\frac{C^{1/2}}{C-1})$. The rest part is the same as in Theorem \ref{th:scs-compl}. So, we obtain the required complexity. The second claim follows from $O(n^{2}\cdot\frac{C^{1/2}}{C-1})=O(n^{2})=O(n^{3}1.728^n)$ if $C=const$.
\Endproof

\section{The Text Assembling Problem}\label{sec:tao}
The algorithm is a modification of the algorithm from \cite{kr2021a}. Our algorithm has better complexity compared to the existing one. Here we present an almost full description of the algorithm for completeness of the presentation.

In this section, we use a quantum subroutine for comparing two strings $u$ and $v$ in the lexicographical order. Let us denote it $\textsc{QCompare}(u,v)$. It is based on the First One search algorithm \cite{dhhm2006,k2014,ll2015,ll2016,kkmsy2022} that is a modification of Grover's search algorithm \cite{g96,bbht98}. The procedure in different forms was discussed in several papers \cite{bjdb2017,ki2019,a2019,kiv2022,kkmsy2022,kl2020,l2020}.
The main property of the subroutine is presented in the following lemma.
\begin{lemma}[\cite{kiv2022}]\label{lm:str-compr}The quantum algorithm $\textsc{QCompare}(u,v)$ compares strings $u$ and $v$ in the lexicographical order in $O(\sqrt{\min{(|u|,|v|)}})$ running time, and the error probability is at most $0.1$.
\end{lemma}

Let us present a quantum algorithm for the $\texttt{TAO}(t,S)$ problem. 
Let $long_i$ be an index of the longest string from $S$ that starts in the position $i$ of the string $t$, where $1\leq i \leq m$. Formally, $long_i=j$ if $s^j$ is the longest string from $S$ such that $t[i,i+|s^j|-1]=s^j$. Let $long_i=-1$ if there is no such string $s^j$.
If we construct the array $long=(long_1,\dots,long_m)$, then we can construct $Q=(q_1,\dots,q_r)$ and $I = (i_1,\dots,i_r)$ in $O(m)$ running time. Note that $Q$ and $I$ arrays are a solution to the problem $\texttt{TAO}(t,S)$. A procedure $\textsc{ConstructQI}(long)$ that constructs $Q$ and $I$ by $long$ is presented in Appendix B. If there is no a $(Q,I)$ decomposition of $t$, then the procedure returns $NULL$.
Let us discuss how to construct the array $long$. 

{\bf Step 1.} We start from constructing a suffix array $suf$ by the string $t$. After that,  we present an array $a=(a_1,\dots,a_m)$ such that $a_i=(len_i,ind_i)$, and $a_i$ corresponds to $suf_i$. We compute values $len_i$ and $ind_i$ on the next steps. Here $len_i$ is the length of the longest string $s^j$ that is a prefix of the suffix $t[suf_i,n]$ and $ind_i$ is its index. Before processing all strings we initialize the values $a_i$ by $(0,-1)$ that are neutral values for our future operations.

{\bf Step 2.} We construct a segment tree $st$ for the array $a$ such that a node of the tree stores the maximum of $len_i$ for $i$ belonging to the node's segment.

{\bf Step 3.} We process $s^j$, for each $j\in\{1,\dots, n\}$. We compute the minimal index $low$ and the maximal index $high$ such that each suffix $t[suf_i,n]$ have $s^j$ as a prefix for $low\leq i \leq high$. 
Suffixes in the suffix array are sorted, therefore, all target suffixes are situated sequentially. Hence, we can use the Binary search algorithm for computing  $low$ and $high$. 
The $\textsc{QCompare}$ subroutine from Lemma \ref{lm:str-compr} is used as a string comparator. We present the implementation of the step in Appendix C as $\textsc{SearchSegment}(s^j)$ subroutine. It returns the pair $(low,high)$, or $(NULL,NULL)$ if no suffix of $t$ contains $s^j$ as a prefix.

{\bf Step 4.} We update nodes of the segment tree that correspond to elements of $a$ in range $[low,high]$ by a pair $(|s^j|,j)$.

We repeat Steps 3 and 4 for each string from $S$. After that, we finish the construction of $a$ by application of push operation for the segment tree.

{\bf Step 5.} We can construct $long$ by $a$ and $suf$. 
If $a_i=(len_i,ind_i)$, then we assign $long_{suf_i}\gets ind_i$. We can do it because of the definitions of $long$, $suf_i$, $len_i$, and $ind_i$.

The whole algorithm is presented as Algorithm \ref{alg:quantum}, and its complexity is discussed in Theorem \ref{th:tao}.

\begin{algorithm}[ht]
\caption{The quantum algorithm for the text $t$ constructing from a dictionary $S$ problem}\label{alg:quantum}
\begin{algorithmic}
\State $suf\gets\textsc{ConstructSuffixArray}(t)$
\State $a\gets[(0,-1),\dots,(0,-1)]$\Comment{Initialization by $0$-array}
\State $st \gets\textsc{ConstructSegmentTree}(a)$
\For{$j\in\{1,\dots,m\}$}
\State $(low,high)\gets \textsc{SearchSegment}(s^j)$
\State $\textsc{Update}(st,low,high,(|s^j|,j))$
\EndFor
\State $\textsc{Push}(st)$
\For{$i\in\{1,\dots,n\}$}
\State $(len,ind)\gets \textsc{Request}(st,i)$
\State $long_{suf_i}\gets ind$
\EndFor
\State $(Q,I)\gets\textsc{ConstructQI}(long)$
\State\Return $(Q,I)$
\end{algorithmic}
\end{algorithm}

\begin{theorem}\label{th:tao}
Algorithm \ref{alg:quantum} solves $\texttt{TAO}(t,S)$ problem with  $O\left(m +\log m\cdot\sqrt{nL}\right)$ running time and the error probability at most $1/3$. 
\end{theorem}
\Beginproof
The correctness of the algorithm follows from the construction.
Due to results from Section \ref{sec:tools}, the running time for each of procedures $\textsc{ConstructSuffixArray}$, $\textsc{ConstructSegmentTree}$, and $\textsc{Push}$ are $O(m)$.
Construction of the array $long$, and construction of $\textsc{ConstructQI}$ have $O(m)$ running time because each of them contains just one linear loop.

Due to Lemma \ref{lm:str-compr}, the running time of $\textsc{QCompare}$ for $s^j$ is $O(\sqrt{|s^j|})$. The procedure $\textsc{QSearchSegment}$ invokes $\textsc{QCompare}$ procedure $O(\log m)$ times for each string $s^1,\dots, s^n$. So, the complexity of processing all strings from $S$ is
$O\left(\log m\cdot \sum_{j=1}^n\sqrt{|s^j|}\right)$.

Let us use the Cauchy-Bunyakovsky-Schwarz inequality and $L=\sum_{j=1}^n|s^j|$ equality for simplifying the statement.

\[\leq O\left(\log m\cdot \sqrt{n\sum\limits_{j=1}^n|s^j|}\right)=O\left(\log m\cdot \sqrt{n\cdot L}\right).\]

The total running time is 

\[O\left(m+m+m +m+m +\log m\cdot \sqrt{n L}\right)=
O\left(m +\log m\cdot \sqrt{n L}\right)\]

Let us discuss the error probability. Only the $\textsc{QCompare}$ subroutine can have an error. We invoke it several times. The subroutine's error probability is at most $0.1$. The error can accumulate and reach a probability close to $1$. At the same time, $\textsc{QCompare}$ is the First One Search algorithm for a specific search function. Due to \cite{k2014,lmrss2011}, we can modify the sequential invocations of the First One Search algorithm to an algorithm that has the error probability at most $1/3$ without complexity changing. 
\Endproof
\section{Conclusion}\label{sec:conclusion}
We present a quantum algorithm for the SSP or SCS problem.  It works faster than existing classical algorithms. At the same time, there are faster classical algorithms in the case of restricted length of strings \cite{gkm2013,gkm2014}. It is interesting to explore quantum algorithms for such a restriction. Can quantum algorithms be better than classical counterparts in this case?
Another open question is approximating algorithms for the problem. As we mentioned before, such algorithms are more useful in practice. So, it is interesting to investigate quantum algorithms that can be applied to practical cases.

In the case of the Text Assemble problem, upper and lower bounds are far apart. It is interesting to find a better quantum lower bound or improve the upper bound.

For both problems an open question is developing a quantum algorithm for the case with possible typos.

\textbf{Acknowledgments}
\noindent
This work is supported by National Natural Science Foundation of China under Grant No. 61877054, 12031004,12271474, and the Foreign Experts in Culture and Education Foundation under Grant No. DL2022147005L. Kamil Khadiev thanks these projects for supporting his visit.
A part of the research was supported by the subsidy allocated to Kazan Federal University for the state assignment in the sphere of scientific activities, project No. 0671-2020-0065. 


\textbf{References}

\bibliographystyle{plain}
\bibliography{report}

\newpage
\appendix{ An Implementation of Segment Tree's Operations}
Assume that the following elements are associated with each node $v$ of the segment tree:
\begin{itemize}
\item $g(v)$ is the target value for a segment tree. If it is not assigned, then $g(v)=-\infty$
\item $d(v)$ is an additional value. 
\item $l(v)$ is the left border of the segment that is associated with the node $v$.
\item $r(v)$ is the right border of the segment that is associated with the node $v$.
\item $LeftC(v)$ is the left child of the node.
\item $RightC(v)$ is the right child of the node
\end{itemize}
If a node $v$ is a leaf, then $LeftC(v)=RightC(v)=NULL$.
Let $st$ be associated with the root of the tree.

Let us present Algorithm \ref{alg:update} for the Update operation. It is a recursive procedure.

\begin{algorithm}[ht]
\caption{$\textsc{Update}(v,i,j,x,y)$. Update $[i,j]$ segment by $(x,y)$ for a segment tree with a root $v$}\label{alg:update}
\begin{algorithmic}
\If{$l(v)=i$ and $r(v)=j$ and $g(v)<x$}
\State $g(v)\gets x$, $d(v)\gets y$
\Else
\State $m\gets r(LeftC(v))$
\If{$m\geq j$}
\State $\textsc{Update}(LeftC(v),i,j,x,y)$
\Else
\If{$m<i$}
\State $\textsc{Update}(RightC(v),i,j,x,y)$
\Else
\State $\textsc{Update}(LeftC(v),i,m,x,y)$
\State $\textsc{Update}(RightC(v),m+1,j,x,y)$
\EndIf
\EndIf
\EndIf
\end{algorithmic}
\end{algorithm}

Let us present the implementation for the Push operation in Algorithm \ref{alg:push} and \ref{alg:push-base}. It is a recursive procedure.
\begin{algorithm}[ht]
\caption{$\textsc{Push}(v)$. Push operation a segment tree with a root $v$}\label{alg:push}
\begin{algorithmic}
\State $\textsc{Push\_Base}(v, -1,-1)$
\end{algorithmic}
\end{algorithm}
\begin{algorithm}[ht]
\caption{$\textsc{Push\_Base}(v, g,d)$. Push operation for a segment tree with a root $v$ and assigning a values $g$ and $d$}\label{alg:push-base}
\begin{algorithmic}
\If{$v\neq NULL$}
\If{$g(v)>g$}
\State $g\gets g(v)$, $d\gets d(v)$
\EndIf
\State $\textsc{Push\_Base}(LeftC(v),g,d)$
\State $\textsc{Push\_Base}(RightC(v),g,d)$
\EndIf
\end{algorithmic}
\end{algorithm}

\appendix{ The Implementation of $\textsc{ConstructQI}(long)$ Procedure for $\texttt{TAO}(S,t)$ Problem}
Algorithm \ref{alg:qi} contains the implementation.
\begin{algorithm}[ht]
\caption{$\textsc{ConstructQI}(long)$. Constructing $Q$ and $I$ from $long$}\label{alg:qi}
\begin{algorithmic}
\State $d \gets 1$
\State $i_d\gets long_1$
\State $q_d\gets 1$
\State $left\gets 2$
\State $right\gets |s^{i_1}|+1$
\While{$q_d<n$}
\State{$max\_i\gets left$}
\State{$max\_q\gets -1$}
\If{$long_{left}>0$}
\State{$max\_q\gets left + |s^{long_{left}}|-1$}
\EndIf
\For{$j\in \{left+1,\dots,right\}$}
\If{$long_j>0$ and $j + |s^{long_j}|-1>max\_q$}
\State{$max\_i\gets j$}
\State{$max\_q\gets j + |s^{long_j}|-1$}
\EndIf
\EndFor
\If{$max\_q= -1$ or $max\_q< right$}
\State Break the While loop and \Return $NULL$ \Comment{We cannot construct another part of the string $t$}
\EndIf 
\State $d\gets d+1$
\State $i_d\gets long_{max\_i}$
\State $q_d\gets max\_i$
\State $left\gets right+1$
\State $right\gets max\_q+1$
\EndWhile
\State \Return $(Q,I)$
\end{algorithmic}
\end{algorithm}

\appendix{ The Implementation of $\textsc{SearchSegment}(u)$ Procedure for $\texttt{TAO}(S)$ Problem}
Algorithm \ref{alg:search} contains the implementation.

\begin{algorithm}[ht]
\caption{$\textsc{SearchSegment}(u)$. Searching for an indexes segment  of suffixes for $t$ that have $u$ as a prefix}\label{alg:search}
\begin{algorithmic}
\State $low\gets NULL$, $high\gets NULL$
\State $l\gets 1$
\State $r\gets n$
\State $Found\gets False$
\While{$Found=False$ and $l\leq r$}
\State $mid\gets (l+r)/2$
\State $pref\gets t[suf_{mid},\min(n,suf_{mid}+|u|-1)]$
\State $pref1\gets t[suf_{mid-1},\min(n,suf_{mid-1}+|u|-1)]$
\State $compareRes \gets \textsc{QCompare}(pref,u), compareRes1 \gets \textsc{QCompare}(pref1,u)$
\If {$compareRes=0$ and $compareRes1=-1$}
\State $Found\gets true$
\State $low\gets mid$
\EndIf
\If {$compareRes< 0$}
\State $l\gets mid+1$
\EndIf
\If {$compareRes\geq 0$}
\State $r\gets mid-1$
\EndIf
\EndWhile
\If{$Found=True$}
\State $l\gets 1$
\State $r\gets n$
\State $Found\gets False$
\While{$Found=False$ and $l\leq r$}
\State $mid\gets (l+r)/2$
\State $pref\gets t[suf_{mid},\min(n,suf_{mid}+|u|-1)]$
\State $pref1\gets t[suf_{mid+1},\min(n,suf_{mid+1}+|u|-1)]$
\State $compareRes \gets \textsc{QCompare}(pref,u), compareRes1 \gets \textsc{QCompare}(pref1,u)$
\If {$compareRes=0$ and $compareRes1=+1$}
\State $Found\gets true$
\State $high\gets mid$
\EndIf
\If {$compareRes\leq 0$}
\State $l\gets mid+1$
\EndIf
\If {$compareRes> 0$}
\State $r\gets mid-1$
\EndIf
\EndWhile
\EndIf
\State \Return $(low, high)$
\end{algorithmic}
\end{algorithm}

\appendix{ The Proof of Lemma \ref{lm:scs-graph-connection}}\label{apx:scs-graph}

\textbf{Lemma \ref{lm:scs-graph-connection}} \textit{
The path $P$ that visits all the nodes of $G$ exactly once and has the maximal possible weight corresponds to the shortest common superstring $t$ for the sequence $S$.
}

\Beginproof
Firstly, let us show that a path that visits all nodes of $G$ corresponds to a superstring. 
Assume that we have a path $P$. We collect a string $t$ by $P$ according to \textsc{ConstructSuperstringByPath} procedure. So, we add a string corresponding to each node at least once. Therefore, $t$ contains all strings at least once.

Secondly, let us compute the length of the string $t$ collected by $P=(v^{i_1},\dots,v^{i_\ell})$. The first node from the string adds $|s^{i_1}|$ symbols to $t$. Each other node $v^{i_j}$ adds $|s^{i_j}|-w(i_{j-1},i_j)$ symbols. Therefore, the total length is 
\[|s^{i_1}|+|s^{i_2}|-w(i_{1},i_2)+\dots+|s^{i_\ell}|-w(i_{\ell-1},i_\ell)=\sum_{j=1}^\ell|s^{i_j}|-\sum_{j=2}^\ell w(i_{j-1},i_j)=\sum_{j=1}^\ell|s^{i_j}|-w(P).\]
Note that if $P$ visits each node exactly once, then $\sum_{j=1}^\ell|s^{i_j}|=\sum_{j=1}^n|s^{j}|=L$, and the length of the string $t$ is $L-w(P)$.

Remind that according to the \textsc{ConstructTheGraph} procedure, the graph $G$ is a full graph.
Let $P'=(v^{i_1},\dots v^{i_\ell})$ be a path that visits a node $v^{i_r}$ at least twice. In that case, we remove the node $v^{i_r}$ from the path. The new path $P''=(v^{i_1},\dots, v^{i_{r-1}}, v^{i_{r+1}},\dots, v^{i_\ell})$ still valid because the graph is full and any pair of nodes are connected. Let us compare lengths $m'$ and $m''$ of two superstrings that are collected by $P'$ and $P''$, respectively. 

The connection between $v^{i_{r-1}}$ and $v^{i_{r+1}}$ in $P''$ gives us a string $u''$ which a prefix is $s^{i_{r-1}}$ and a suffix is $s^{i_{r+1}}$. The length of $u''$ is minimal possible because $w(i_{r-1},i_{r+1})$ is the maximal overlap. The connection between $v^{i_{r-1}}$ and $v^{i_{r+1}}$ via $v^{i_{r}}$  in $P'$ gives us a string $u'$ which a prefix is $s^{i_{r-1}}$ and a suffix is $s^{i_{r+1}}$. Definitely, $|u'|\geq|u''|$. Therefore, $m'\geq m''$.
 So, superstrings that are collected by paths visiting each node exactly once are shorter than paths that visit some node at least twice. 

Thirdly, let us show that any path $P''$ that does not visit all nodes and corresponds to a superstring can be extended to a path that visits all nodes such that the new collected superstring is not longer than the original one.
Assume that we have a path $P''=(v^{i_1},\dots,v^{i_\ell})$ that does not visit $v^r$, end the corresponding string $t$ is a superstring. Therefore, $t$ contains $s^r$ as a substring too. So, there is $v^{i_j}$ and $v^{i_{j+1}}$ such that the string $u$ that is collected from $s^{i_j}$ and $s^{i_{j+1}}$ by removing their overlapping has  $s^r$ as a substring. Note that it cannot be three sequential nodes from the path $P''$ because otherwise, the middle string should be a substring of $s^r$. At the same time, it is an impossible situation because we remove all duplicates and substrings in the first step of our algorithm.
%
Assume that $s^r$ starts in $u$ in position $j_s$ and finishes in position $j_f$. Therefore, $u[1,j_f]$ has $s^{i_j}$ as a prefix and $s^r$  as a suffix. At the same time, $u[j_s,|u|]$ has  $s^r$ as a prefix and $s^{i_{j+1}}$ as a suffix.

Let us update $P''$ by inserting $v^r$ between $v^{i_j}$ and $v_{i_j+1}$. $P'''=(v^{i_1},\dots,v^{i_j}, v^r,v_{i_j+1},\dots,v^{i_\ell})$.
Let us look to the string $u'$ that is collected from  $s^{i_j}$, $s^{r}$, and  $s^{i_{j+1}}$ according to connection of $v^{i_j}, v^r,$ and $v_{i_j+1}$ in $P'''$.
Assume that $s^r$ starts in $u'$ in position $j'_s$ and finishes in position $j'_f$. Therefore, $u'[1,j'_f]$ has $s^{i'_j}$ as a prefix and $s^r$  as a suffix; and it is the shortest possible such string. At the same time, $u'[j'_s,|u'|]$ has  $s^r$ as a prefix and $s^{i_{j+1}}$ as a suffix; and it is the shortest possible such string.

Hence, $|u'[1,j'_f]|\leq |u[1,j_f]|$, and $|u'[j'_s,|u'|]|\leq |u[j_s,|u|]|$.
At the same time,
\[|u|=|u[1,j_f]|+|u[j_s,|u|]|-|s^r|\geq |u'[1,j'_f]|+|u'[j'_s,|u|]|-|s^r|=|u'|\]

We can see that other parts of $P'''$ and $P''$ are the same. Therefore, strings $t'''$ and $t''$ collected by these paths are such that $|t'''|\leq |t''|$. Hence, any path that does not visit all nodes and corresponds to a superstring can be completed by the rest nodes and the new corresponding superstring does not become longer. Therefore, we can search the required path only among paths that visit all nodes exactly once.

Finally, let us show that if $P$ has the maximal weight and visits each node exactly once, then it is the shortest superstring.
Assume that we have another path $P'$ that visits each node exactly once but the weight $w(P')<w(P)$.
Let $m$ and $m'$ be the lengths of the superstrings collected by $P$ and $P'$, respectively.
Then, $m'=L-w(P')>L-w(P)=m'$. Therefore, the superstring collected by $P$ is the shortest superstring.
\Endproof


\end{document}